\documentclass[aps,pra,twocolumn,twoside,floatfix,superscriptaddress,showpacs]{revtex4}
\usepackage{hyperref}
\usepackage{bbm}
\usepackage{amsmath, amssymb}
\usepackage{color}
\usepackage{graphicx,epsfig}
\usepackage{pifont}

\def\C{\mbox{\bf C}}

\begin{document}

\title{Quantum entanglement of nanocantilevers}
\author{C. Joshi}
\email{cj53@hw.ac.uk}
\affiliation{SUPA, Department of Physics, Heriot-Watt University,
Edinburgh, EH14 4AS, UK}
\author{A. Hutter}
\affiliation{SUPA, Department of Physics, Heriot-Watt University,
Edinburgh, EH14 4AS, UK}
\affiliation{Department of Physics and Astronomy, University of W\"urzburg, D-97074 W\"urzburg, Germany}
\author{F. E. Zimmer}
\affiliation{SUPA, Department of Physics, Heriot-Watt University, 
Edinburgh, EH14 4AS, UK}
\affiliation{Max Planck Institute for the Physics of Complex Systems, D-01187 Dresden, Germany}
\author{M. Jonson}
\affiliation{SUPA, Department of Physics, Heriot-Watt University, Edinburgh, EH14 4AS, UK}
\affiliation{Department of Physics, University of Gothenburg, SE-412 96 G{\"o}teborg, Sweden}
\affiliation{Department of Physics, Division of Quantum Phases \& Devices, Konkuk University, Seoul 143-701, Korea}
\author{E. Andersson}
\affiliation{SUPA, Department of Physics, Heriot-Watt University,
Edinburgh, EH14 4AS, UK}
\author{P. \"Ohberg}
\affiliation{SUPA, Department of Physics, Heriot-Watt University,
Edinburgh, EH14 4AS, UK}

\begin{abstract}

We propose a scheme to  entangle two mechanical 
nanocantilevers through indirect interactions mediated by a 
gas of ultra cold atoms. We envisage a system of 
nanocantilevers  magnetically coupled to a Bose-Einstein 
condensate of  atoms and focus on studying the dark states 
of the system. These dark states are entangled states of the 
two nanocantilevers, with no coupling to the atomic condensate. 
In the absence of dissipation, the degree of entanglement is 
found to oscillate with time, while if dissipation is included, the 
system is found to relax to a statistical mixture of dark states 
which remains time independent until the inevitable thermal 
dephasing destroys the nanocantilever coherence. 
This opens up the possibility of achieving long-lived entangled 
nanocantilever states.
\end{abstract}

\maketitle

\section{Introduction}
There have been long-standing arguments regarding the validity 
of quantum mechanics in the macroscopic world. The last two 
decades have seen an unprecedented rise of interest in studying 
the crossover between quantum and classical mechanics. These 
studies, both theoretical and experimental, have the potential to 
improve our understanding of decoherence, which is assumed 
to degrade any quantum system to its classical counterpart. Over 
the past few years, there have been many fascinating experiments 
to prepare and detect quantum states of macroscopic objects 
\cite{mbru}.  In this quest for studying the level of ``quantumness" 
present in macroscopic objects, tremendous progress has been 
achieved in exploring the quantum regime of micro- and 
nano-mechanical systems \cite{schwab}. These offer a very 
promising playground for studying the quantum-classical 
crossover. Miniaturized cantilevers contain a macroscopic 
number of atoms and can be fabricated to have very high 
resonant frequencies and exceedingly large quality factors 
\cite{roch},  thereby guarding against  the effects of decoherence. 

The advancement in techniques such as laser cooling of mechanical 
resonators \cite{choh,iwra,simo} has brought quantum state 
preparation within experimental reach. Experimental advances have 
also fueled a surge of interest in coupling these condensed-matter 
systems to other quantum systems with well-understood quantum 
properties, such as Bose-Einstein condensates or electron spins. 
Such hybrid quantum systems are promising candidates for 
ultra-precise measurements \cite{yjwa}. Very recently, O'Connell et al. 
\cite{ocon} were able to cryogenically cool a mechanical resonator to 
its quantum ground state, and were also successful in strongly 
coupling it to a superconducting qubit to read out the motion of the 
resonator. This success heralds a new era in investigating the quantum 
behavior of nanomechanical systems.

The study of entanglement and superpositions of macroscopic objects 
is of prime interest. These phenomena have been experimentally 
demonstrated for mesoscopic systems, by interferometry of molecules 
\cite{marn, lhac} and entangling of atomic ensembles \cite{bjul}.  There 
are proposals to generate spatially separated superposition states of 
nanomechanical systems using controlled interaction through a 
Cooper-pair box \cite{adar}, and to use nanomechanical systems as a 
quantum bus for quantum information processing \cite{ancl, ltia}.  Bose 
and Agarwal have proposed a scheme to entangle two nanocantilevers through 
a Cooper-pair box \cite{bose}, while Treutlein et al. have presented a 
scheme to entangle the vibrational mode of a  nanocantilever with the 
spin degree of freedom of an ultracold Bose-Einstein condensate (BEC) 
\cite{ptre}. Here, a nanocantilever is coupled to a cloud of ultracold 
atoms through an external magnet. This scheme is interesting since it 
would allow us to couple  two nanocantilevers to the same cloud of 
ultracold atoms, through which the nanocantilevers can interact with 
each other. Hunger et al. were recently able to resonantly couple the 
vibrations of a micromechanical oscillator to the motional degree of 
freedom of an ultracold BEC in a trap using the atom-surface interactions 
\cite{davi}.  In their setup, the ultracold BEC was trapped at  a few 
${\rm \mu m}$ from the surface of the cantilever.  When the cantilever 
vibrates, the surface potential becomes time dependent and modulates 
the trapping potential for the ultracold BEC. This in turn excites the 
atomic motion and generates coupling between the two.

Motivated by recent experimental results, we here theoretically investigate 
the  possibility of entangling two nanocantilevers coupled via an ultracold 
Bose gas. We model the nanocantilevers as quantum harmonic oscillators 
which have been cooled near to their ground states, so that the average 
number of excitations is much smaller than unity. The ultracold Bose gas is 
assumed to comprise $N$ two-level atoms, and is modeled using the Dicke 
formalism \cite{dicke}.  We study the time evolution with and without 
dissipation, and find that entanglement arises in either case. Unitary evolution 
of the system leads to generation of time-varying entanglement between the 
two nanocantilevers. The dynamics becomes increasingly  complex the more 
highly excited the studied system is. In the presence of dissipation in the 
atomic gas, the two nanocantilevers will be entangled in  a steady state, which 
is decoupled from the atomic gas and has a life time that is only limited by the 
inevitable dephasing caused by the environment. This opens the possibility of 
achieving long lived ``macroscopic'' entangled states. 
 
The paper is organized in the following way. Section \ref{sec:model} introduces 
the theoretical model and the physical setup for a system of two nanocantilevers 
interacting via an ultracold Bose gas. This forms the framework for Sec.
\ref{sec:unitary} in which the unitary evolution of the system is studied using the 
Schr\"odinger and Heisenberg time evolution in different excitation manifolds. 
Decoherence effects are included in the description in section \ref{sec:dissipation} 
while the entanglement present in the system is quantified in section
\ref{sec:entanglement}.  Finally, a discussion concludes the paper in section 
\ref{sec:conclude}.

\section{Theoretical model and physical setup}
\label{sec:model}
We  propose a  scheme to generate entangled states of  two nanocantilevers 
through indirect interactions mediated via an ultracold Bose gas. Let us 
assume that two identical nanocantilevers, each mechanically vibrating in its 
fundamental flexural mode \cite{klek}, have been precooled near to their 
ground states. We model the two nanocantilevers as quantum harmonic 
oscillators, with $\hat{a}^{\dagger}(\hat{a})$ and $\hat{b}^{\dagger}(\hat{b})$ 
as the creation (annihilation) operators for vibron excitations in cantilever $a$ 
and $b$, respectively. The ultracold gas is described as a collection of $N$ 
two-level atoms, modeled by the Dicke formalism \cite{dicke} with 
$\hat{J}_{+}$ ($\hat{J}_{-}$) as the collective spin raising (lowering) operator.  

The physical setup is shown in fig.~\ref{fig0}. The two identical nanocantilevers, 
separated by a distance $2d$ along the $y$-axis, are assumed to be  fabricated  on an atom 
chip  with strong ferromagnets attached to their tips.  Equidistant from the tips, 
at a distance $d$, an ultracold Bose gas is confined in a microtrap. The 
magnetic moment {\boldmath $\mu$} of each ferromagnet  is pointing in the  
$x$-direction and the ultracold atoms are subject to a stationary magnetic field 
of strength $B_{0}$ along the $z$-direction, which becomes the quantization 
axis for the ultracold atoms. For the purpose of providing an identical  trapping 
potential for all the hyperfine states of the atoms \cite{ptre} we envisage an 
optical or electrodynamic trap \cite{altgeomtry}. In such a trap all the atoms in 
the BEC couple simultaneously to the quantized motion of the resonator.

Under the dipole approximation, the  $x$-component of the magnetic field  at 
the center of the trap, produced by the ferromagnet  on the tip of  cantilever 
$a$, is 
\begin{equation}
B_{x}=-\frac{\mu_{0}\mu}{4 \pi y^{3}}=-\frac{\mu_{0}\mu}{4 \pi [d+y_{a}(t)]^{3}}\,,
\end{equation}
where $\mu_0$ is the vacuum permeability and $y_{a}(t)$ is the time 
dependent deflection of the tip of nanocantilever $a$. For small 
displacements of the  cantilever this expression can be expanded to linear 
order in $y_{a}(t)$, so that 
\begin{equation}
B_{x}=-\frac{\mu_{0}\mu}{4 \pi d^{3}}\left[1-3\frac{y_{a}(t)}{d}\right].
\end{equation}
Thus $y_{a}(t)$ transduces the vibrational motion of cantilever $a$ to an 
oscillating magnetic field given by
\begin{equation}
{\bf B}=G_{m}y_{a}(t) \hat{x}
\end{equation}
at the location of the ultracold atoms, where $G_{m}$=$3\mu \mu_{0}/4 \pi d^{4}$ 
is the magnitude of the magnetic field gradient in the $y$-direction. By making 
use of, e.g., the quadratic Zeeman effect selected hyperfine states of the atoms 
in the ultracold gas can be decoupled from other states. Transitions from  
$|F=1,m=0\rangle  \equiv |0\rangle$ to $|F=1,m=-1\rangle \equiv |1\rangle$ can, 
for instance, be resonantly coupled to the quantized bending motion of the 
cantilever by tuning the Larmor frequency. In this way coupling is achieved 
between the fundamental bending mode of cantilever $a$ and the collective 
spin of the  ultracold gas through the Zeeman interaction,
\begin{equation}
-{\boldmath \mu}_{atom} \cdot  {\bf B(t)}=\hbar \kappa_{a} \frac{1}{\sqrt{N}}(\hat{J}_{+}+\hat{J}_{-}) (\hat{a}+\hat{a}^{\dagger}),
\end{equation}
where {\boldmath $\mu$ }$_{atom}$ is the magnetic moment of the ultracold gas and 
$\kappa_{a}$=$\mu_{B}G_{m}\sqrt{N}a_{0}/\sqrt{8}\hbar$ is the coupling 
constant for the collective atomic spin coupled to a single vibrational mode 
of cantilever $a$. Here $N$ is the number of atoms in the gas and 
$a_{0}=\sqrt{\hbar/2m_{\it eff}\omega_{0}}$ is the amplitude of the zero-point 
fluctuations of  cantilever $a$, which has  effective mass $m_{\it eff}$ and 
angular oscillation frequency $\omega_{0}$. The coupling between cantilever 
$b$ and the ultracold Bose gas is described by an analogous term. 

In the regime of strong coupling between the collective spin of the ultracold 
gas and the vibrational motion of the nanocantilevers, the system is analogous 
to a conventional cavity-QED system. For a 
nanocantilever with mass $m \sim 10^{-16}$~{\rm kg  and resonance frequency 
$\omega_{0}/2 \pi \sim$1~$\rm{MHz}$, which is separated from an atomic cloud 
of $N \sim 100$ atoms by a distance of $d\sim 250$~{\rm nm},  the interaction 
produced by a small  \cite{FiniteMagnetRemark} disk-shaped magnet containing 
$N_{mag}\sim 10^{6}$ nickel atoms corresponds to a coupling constant 
$\kappa_{a} \sim 100~{\rm Hz}$.

In the particular geometry we envisage for physically entangling the two 
nanomechanical systems, direct interaction between the magnetic dipoles 
on the two cantilever tips can be neglected. To see this, one can compare the 
two interaction strengths, one due to the direct interaction between the magnetic 
cantilever tips and the other due to the interaction between either cantilever 
magnet and the collective spin of the ultracold gas. The ratio between these 
interaction terms is
 \begin{equation}
 \frac{H_{~Zeeman}}{H_{~direct}} \approx 4\frac{d \sqrt{N}}{a_{0}N_{mag}}  \sim 25 \, .
\end{equation}
However, even if this direct interaction is included, the qualitative behavior of 
the system does not change. We must stress, though, that the role of the BEC
is essential for generating a time independent statistical mixture of (dark) 
entangled states. In the absence of atoms a pair of nanocantilevers 
interacting via the dipole-dipole interaction may well exhibit time dependent 
entanglement. It would, however, be difficult to use or capture these states 
because one would have to, for instance, be able to control the direct 
interaction between the cantilevers such that it could be switched off at a 
chosen time.

\begin{figure}[t]
   \begin{center}
	\includegraphics[width=0.5\textwidth]{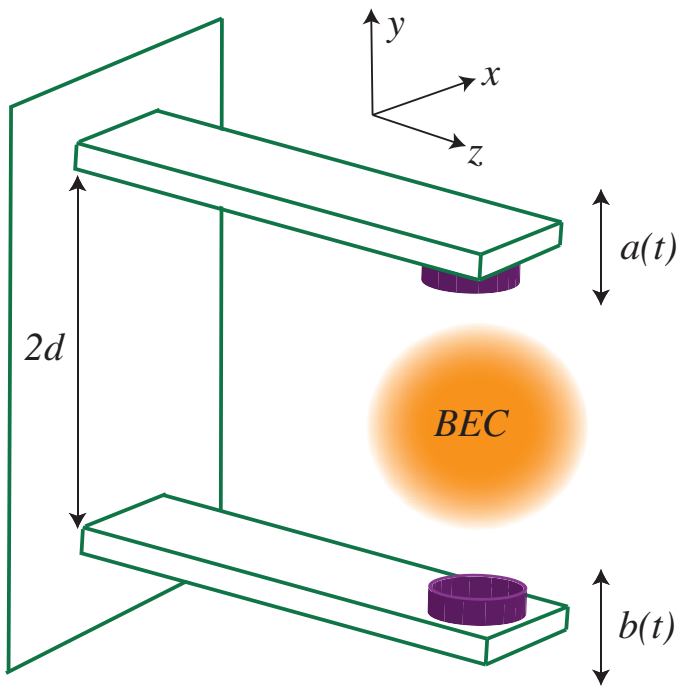}
        \caption{\label{fig0} (Color online) Physical setup for the proposed scheme for entangling 
        two nanocantilevers. Two identical nanocantilevers have  strong ferromagnets 
        attached to their tips. The cantilevers are placed  equidistant from an ultracold 
        gas of atoms, which is  confined to a microtrap. Each ferromagnet couples the 
        vibrational motion of a nanocantilever to the collective spin $J$ of the ultracold gas. }
   \end{center}
\end{figure}

If we assume \cite{NoteBroadening} that the cloud of ultracold atoms can be 
represented by a single quantized mode of frequency $\omega_{a}$, which can be 
tuned to resonate with the frequencies $\omega_{0}$ of the quantized nano-resonators, 
then the dynamics of the two nanocantilevers interacting via the ultracold gas is well 
described by the Jaynes-Cummings Hamiltonian \cite{fwcu}. 

With the dipole and rotating wave approximations, and taking $\kappa_a=\kappa_b=\kappa$, the Hamiltonian for our system is
\begin{eqnarray}\label{1}
\frac{ H}{\hbar}&=&\omega_{0}(\hat{J_{z}}+\hat{a}^{\dagger} \hat{a}+\hat{b}^{\dagger} \hat{b})+
[\kappa \frac{1}{\sqrt{N}}(\hat{a} +\hat{b} ) \hat{J}_{+} +h.c.]\\ \nonumber
&=& H_{0}+ H_{I},
\end{eqnarray}
where $ H_{0}=\omega_{0}(\hat{J_{z}}+\hat{a}^{\dagger} \hat{a}+\hat{b}^{\dagger} \hat{b})$ 
and $ H_{I}=\kappa(\hat{a} +\hat{b} ) \hat{J}_{+} +h.c.$ are the free and interaction parts 
of the Hamiltonian respectively, and $J_{z}$ is the $z$ component of the collective 
angular momentum of the ultracold gas. In the Dicke model, the collective spin of the 
ultracold BEC is $J$ = $N/2$, where $N $ is the number of atoms. A general Dicke state 
is represented as $ |M,J\rangle$, where $J+M$ is the  number of excitations and $J-M$ 
is the number of atoms in the ground state \cite{yyam}. In this notation,  $|-J,J\rangle$ 
represents the ground state and $|J,J\rangle$ the highest excited state. Spin raising and 
lowering operators act on a general Dicke state as 
\begin{eqnarray}
\hat{J_{+}}|M,J\rangle&=& \sqrt{J(J+1)-M(M+1)}|M+1,J\rangle\\
\hat{J_{-}}|M,J\rangle&=& \sqrt{J(J+1)-M(M-1)}|M-1,J\rangle.
\end{eqnarray}

After these initial considerations, we proceed to study the unitary evolution of the system 
followed by an investigation of the dissipative dynamics of the system.

\section{Unitary Evolution}
\label{sec:unitary}
\subsection{Schr\"odinger picture}
 \label{sec:unitarysch}
We shall now consider  a unitary evolution of the system of two nanocantilevers interacting 
via an ultracold gas of atoms. As mentioned before, we assume that the nanocantilevers 
have been cooled near to their quantum ground states and  that only the few lowest-energy 
Fock states of the quantized oscillators are populated. In the Schr\"odinger picture, we 
restrict the excitation subspace of the two nanocantilevers to one, two and three quanta of 
vibrational excitations. 

To study the dynamics in the Schr\"odinger picure we shall assume that the ultracold gas  comprises $N$ two-level atoms, and that the maximum number of excitations in the gas is restricted to one. This is a reasonable assumption since we have found that in an $n$-excitation subspace with $n \ll N$, inclusion of higher excitations in the gas changes the Rabi frequency without interfering with the qualitative behavior of the system. With this assumption, there are only two global states of the ultracold BEC, namely $|-J,J\rangle$ and $|-J+1,J\rangle$. The action of $\hat{J}_{+}$ and $\hat{J}_{-}$ on these states is
\begin{eqnarray}
\hat{J_{+}}|-J,J\rangle&=& \sqrt{N}|-J+1,J\rangle\\
\hat{J_{-}}|-J+1,J\rangle&=& \sqrt{N}|-J,J\rangle.
\end{eqnarray}
In the discussion to follow, we
denote the collective ground state $|-J,J\rangle$ of the ultracold gas by $|g\rangle$ and its first excited state $|-J+1,J\rangle$ by $ |e\rangle$. 

To a first approximation, we neglect any dissipation present in the system and use the Schr\"odinger equation to study the unitary dynamics of the system. In the interaction picture, a general initial state of the cantilever-gas system in the  one-excitation manifold can be written as
 \begin{eqnarray}{\label 2}
|\Psi_1(t)\rangle&=&e^{-iH_0t/\hbar}\left[ C_{g,0,1}(t) |g,0,1\rangle \right.
\\
&&+\left.C_{g,1,0}(t)|g,1,0\rangle+C_{e,0,0}(t) |e,0,0\rangle\right],\nonumber
 \end{eqnarray}
where $|g,l,m\rangle$ ($|e,l,m\rangle$) represents a state with the ultracold gas in its ground (excited) state, cantilever $a$ with $l$ excitations and cantilever $b$  with $m$ excitations. Using \eqref{2}, the Schr\"odinger equation becomes 
\begin{eqnarray}{\label 3}
\frac{\partial}{\partial t} \C_1(t)&=&\beta \C_1(t),
\end{eqnarray}
where 
\begin{equation*}
\beta= -i\left[ \begin{array}{ccc} 
 {0} & {0} &\kappa\\ 
 {0} & {0} &\kappa\\  
 \kappa& \kappa&{0}
\end{array}
\right]
\end{equation*}
and 
\begin{equation*}
\C_1(t)= \left[ \begin{array}{c} C_{g,1,0}(t) \\ C_{g,0,1}(t)\\ C_{e,0,0}(t) \end{array} \right].
\end{equation*}
  The coupling matrix $\beta$ has the eigenvalues
$\lambda_0=0$, $\lambda_\pm$=$\pm i\sqrt{2}\kappa$, with the corresponding normalized eigenvectors given by
\begin{eqnarray} 
&|\lambda_0\rangle
=\frac{1}{\sqrt{2}}\left[ \begin{array}{c} \ 1\\  -1\\\ 0 \end{array} \right],~~
|\lambda_{-}\rangle
 = \frac{1}{2}\left[ \begin{array}{c} -1\\ -1 \\ \sqrt{2} \end{array} \right] ,&\\~~
&|\lambda_{+}\rangle
 = \frac{1}{2}\left[ \begin{array}{c} 1\\ 1 \\ \sqrt{2} \end{array} \right].&
 \nonumber
 \end{eqnarray} 
The eigenstate $|\lambda_0\rangle$  has no excitation in the atomic gas}.  Such a state is termed a dark state or trapped state, because  its population does not decay even in the presence of dissipation in the atomic gas. Dark states have been extensively studied in the context of decoherence-free subspaces 
 \cite{mbpl}. There have been many proposals to prepare qubits in such states and hence protect them from decoherence \cite{almu}. The population in states orthogonal to dark states undergo dissipation and ultimately relax to the ground state, while the population of the dark states remains intact.
 
If the two nanocantilevers couple to the ultracold gas with different coupling strengths $\kappa_{a}, \kappa_{b}$, then the dark state in the one-excitation manifold takes the form
\begin{equation}\label{state}
|\lambda_0\rangle
=\frac{1}{\sqrt{(\kappa_{a}^{2}+\kappa_{b}^{2})}}\left(\kappa_a|g,0,1\rangle- \kappa_{b}|g,1,0\rangle \right),
  \end{equation} 
which is a maximally entangled state of the two nanocantilevers for $\kappa_{a}=\kappa_{b}$. 
It is interesting to see from Eq.~(\ref{state}), that by tuning the interaction strengths $\kappa_{a}$ and $\kappa_{b}$, adiabatic state transfer \cite{jore} can be achieved between the two cantilevers. This can be realized, for instance,  by instead using electromagnets, making it possible to vary the coupling strengths between the vibrational mode of each nanocantilever and the spin of the ultracold Bose gas. The fabrication of such electromagnets on nanocantilevers  is obviously a challenging  task, but recent advancements in nanofabrication techniques suggest that it might be possible in the future \cite{electromagnet}.
We will further discuss the relevance of dark states for our scheme in the next section, where we will study the dissipative dynamics of the system.  

\begin{figure}[t]
\begin{center}
\includegraphics[width=0.5\textwidth]{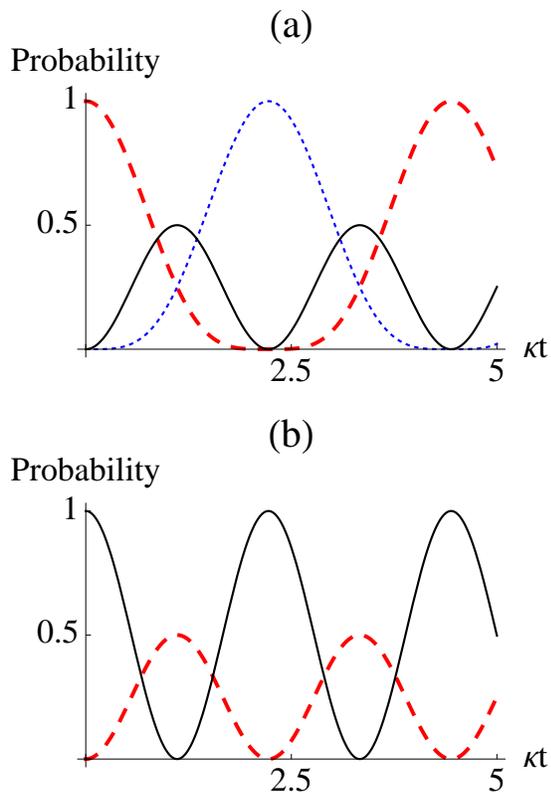}
\caption{\label{fig1} (Color online) 
Time evolution of the occupation probability for
(a) the states $|g,1,0\rangle$ (dashed), $|g,0,1\rangle$ (dotted), and $|e,0,0\rangle$ (solid) with the initial condition
 $C_{g,1,0}(0)=1$,  and (b)  for the states $|g,1,0\rangle$ and $|g,0,1\rangle$ (dashed, identical) and $|e,0,0\rangle$ (solid) with the initial condition
 $C_{e,0,0}(0)=1$.
 }
\end{center}\end{figure}

Equation \eqref{2} can easily be solved, giving the time-evolved wave function in the one-excitation manifold. With $C_{g,1,0}(0)=1$, the system evolves to 
\begin{equation}\label{old}
|\Psi_1(t)\rangle = \frac{1}{2}\left[ 
\begin{array}{c} 1+\cos(\sqrt{2}\kappa t) \\ 
\cos(\sqrt{2}\kappa t)-1\\
-i\sqrt{2}\sin(\sqrt{2}\kappa t)
\end{array} 
\right].
\end{equation} 
The probability for each of the three basis states to be occupied is plotted as a function of time in  fig.~\ref{fig1}. The excitation is reversibly transferred between the cantilevers and the ultracold atomic gas. We shall explicitly quantify the entanglement present in different excitation manifolds in section \ref{sec:entanglement}, but note already here that the states $|g,1,0\rangle$ and $|g,0,1\rangle$  exhibit  time-varying entanglement. It is also worth noting that the amount of entanglement between the two nanocantilevers strongly depends on the initial state of the system. The amount of entanglement generated between the two nanocantilevers is enhanced if the initial excitation lies in the ultracold gas.

\begin{figure}[t]
   \begin{center}
	\includegraphics[width=0.5\textwidth]{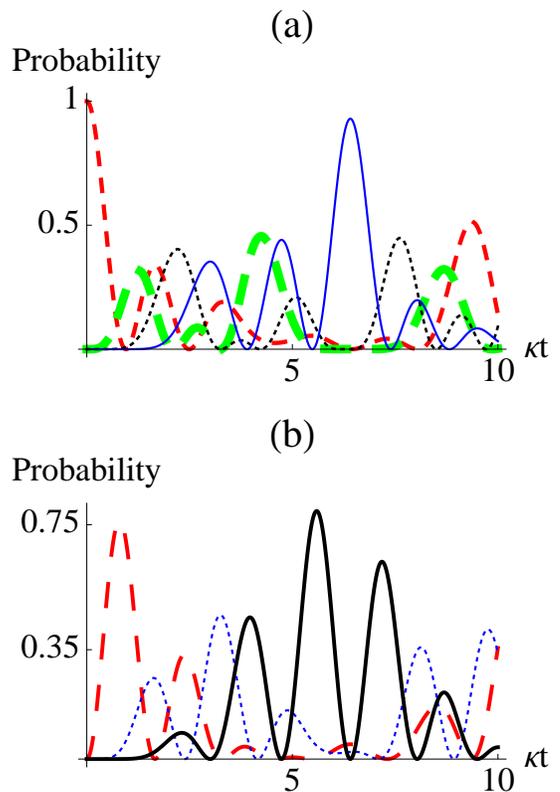}
        \caption{\label{fig3}  (Color online) 
         Occupation probability, as a function of time, for the states (a) $|g,3,0\rangle$ (dashed), $|g,2,1\rangle$ (thick dashed), $|g,1,2\rangle$ (dotted), and $|g,0,3\rangle$ (solid); (b) $|e,2,0\rangle$ (dashed), $|e,1,1\rangle$ (dotted), and $|e,0,2\rangle$ (solid), with the initial condition $C_{g,3,0}(0)=1$.
         }
   \end{center}
\end{figure}

To study the dynamics when
more than one excitation are present in the system, equation \eqref{2} can readily be generalized. 
In the interaction picture, general state vectors for  the system in the  two- and three-excitation manifolds take the form
\begin{eqnarray}
|\Psi_2(t)\rangle&=&e^{-i\frac{H_{0} t}{\hbar}}\left[\sum_{j=0}^2C_{g,j,2-j}(t)|g,j,2-j\rangle \nonumber\right.\\
&&\left.+\sum_{j=0}^1C_{e,j,1-j}(t)|e,j,1-j\rangle\right],
{\label 4}\\
|\Psi_3(t)\rangle&=&e^{-i\frac{H_{0} t}{\hbar}}\left[\sum_{j=0}^3C_{g,j,3-j}(t)|g,j,3-j\rangle \nonumber\right.\\
{\label 5}
&&\left.+\sum_{j=0}^2C_{e,j,2-j}(t)|e,j,2-j\rangle\right].
 \end{eqnarray}
Using \eqref{4} and \eqref{5}, the Schr\"odinger equation reduces to sets of coupled differential equations which can be easily solved.  A typical solution  for a three-excitation subspace is shown in fig.
\ref{fig3}, where occupation probabilities are plotted
as functions of time for the different basis states. The result is a time-varying entanglement between the states in each excitation 
manifold.  It turns out that the system of two nanocantilevers exhibits entanglement in higher excitation subspaces, even for mixed initial states, such as a thermal state.  Thus thermal effects do not necessarily degrade the entanglement of a quantum system.

\subsection{Heisenberg picture} 
The Schr\"odinger and Heisenberg pictures are two equivalent ways of studying the evolution of a quantum system. After studying the system in the Schr\"odinger picture, we now present a compact 
way of studying the unitary evolution of the system in the Heisenberg picture. This method, apart from giving more insight into the dynamics of the system, also allows us to exactly solve the dynamics of the system  
for arbitrarily many excitations.  

Using the Dicke model introduced in section \ref{sec:model}, we have $-J \leq M \leq J$. Since most of the atoms in the  ultracold  gas are  condensed in the ground state, and $N$ is large, we can approximate $[\hat{J_{+}},\hat{J_{-}}]$=2$\hat{J_{z}} \approx-N.$  An equivalent way of invoking this approximation is through the Holstein-Primakoff (HP) transformation \cite{hols}. This
is essentially a mapping from angular momentum spin operators to bosonic operators. Under the HP transformation, the collective angular momentum spin operators transform as 
\begin{eqnarray}
\hat{J}_{+} &=& \hat{c}^{\dagger}\sqrt{N-\hat{c}^{\dagger} \hat{c}}\label{6a}\\
\hat{J}_{-}& =&\hat{c} \sqrt{N-\hat{c}^{\dagger}\hat{c}}\\
\hat{J}_{z}& =& \hat{c}^{\dagger}\hat{c}- {N}/{2},\label{6c}
\end{eqnarray}
where $\hat{c}$ and $\hat{c}^{\dagger}$ are the bosonic  lowering and raising operators which satisfy $[\hat{c},\hat{c}^{\dagger}]=1$. In the large $N$ limit, Eqs. (\ref{6a})-(\ref{6c})  reduce to  $\hat{J}_{+} \approx \hat {c^{\dagger}}\sqrt{N}$,  $\hat{J}_{-} \approx \hat{c}\sqrt{N}$ and  $\hat{J}_{z} \approx \hat{c^{\dagger}} \hat{c}- {N}/{2}$. Under this approximation the Hamiltonian \eqref{1} becomes 
\begin{equation}\label{7}
\frac{ H}{\hbar}=\omega_{0}( \hat{c}^{\dagger} \hat{c}- {N}/{2}+\hat{a}^{\dagger} \hat{a}+\hat{b}^{\dagger} \hat{b})+[\kappa (\hat{a} +\hat{b})\hat{c}^\dagger +h.c.].
\end{equation}

We now introduce  $\hat{p}^{\dagger}_\pm(0)$ and  $\hat{q}^{\dagger}(0)$ as two sets of bosonic creation operators for collective excitations. By choosing
\begin{eqnarray}\label{8}
\hat{p_{\pm}}^\dagger(0)&=&\frac{1}{2}[\hat{a}^\dagger(0) +\hat{b}^\dagger(0)]\pm \frac{1}{\sqrt{2}}\hat{c}^\dagger(0)\\
\hat{q}^{\dagger}(0)&=&\frac{1}{\sqrt{2}}[\hat {a}^\dagger(0)-\hat {b}^\dagger(0)],\label{8}
\end{eqnarray}
the Hamiltonian \eqref{7} becomes diagonal,
\begin{equation}\label{9}
\hat {H}= \sum_{k=\pm} \hbar  \Omega_{k} \hat{ p_{k}} ^\dagger \hat{ p_{k}},
\end{equation}
where $\Omega_{\pm}=\omega_{0} \pm \sqrt{2} \kappa$.  The time-evolved creation operators are 
\begin{eqnarray}
\hat{p_{\pm}}^\dagger(t)&=&e^{i\Omega_{\pm}t}\hat{p_{\pm}}^\dagger(0)\label{10a}\\
\hat{q}^\dagger(t)&=& e^{i\omega_{0} t }\hat{q}^\dagger(0).\label{10b}
\end{eqnarray}
From (\ref{10a}) and (\ref{10b}) we obtain 
\begin{eqnarray}\label{12}
\hat{a}^\dagger(t)&=&\frac{1}{2}[\hat{p_{+}}^\dagger(t)+\hat{p_{-}}^\dagger(t)+\sqrt{2}\hat{q}^\dagger(t)]\\
\hat{b}^\dagger(t)&=&\frac{1}{2}[\hat{p_{+}}^\dagger(t)+\hat{p_{-}}^\dagger(t)-\sqrt{2}\hat{q}^\dagger(t)].
\label{12}
\end{eqnarray}
These  time evolved creation operators $\hat{a}^\dagger(t)$  and $\hat{b}^\dagger(t)$ can conveniently be used to describe the unitary time evolution of the system of two nano-cantilevers and an ultracold gas for arbitrary initial conditions.  Evolution of a state with cantilever $a$ initially in its $n$th excited state can easily be determined from 
$|\Psi_n(t)\rangle=(\hat{a}^\dagger(t))^{n}|g,0,0\rangle$. 
For instance, if $n=1$, we obtain
\begin{equation}
|\Psi_{1}(t)\rangle=\hat{a}^\dagger(t)|g,0,0\rangle
= \frac{1}{2}\left[ 
\begin{array}{c} 1+\cos(\sqrt{2}\kappa t) \\ 
\cos(\sqrt{2}\kappa t)-1\\
i\sqrt{2}\sin(\sqrt{2}\kappa t)
\end{array} 
\right],
\end{equation}  
where the basis vectors are the same as for Eq. \eqref{old}, obtained in the Schr\"odinger picture. The results are identical apart from a phase shift on the last component; this is a general feature of the 
Schr\"odinger vs. the Heisenberg picture.
The dark states of the system are seen to correspond to excitations of the mode labeled $q$, so that a dark state with $n$ excitations is given by $|\Psi_{q,n}(t)\rangle=(\hat{q}^\dagger(t))^{n}|g,0,0\rangle$. This way, dark states corresponding to different numbers of excitations in the system can easily be computed.

Before concluding this section, we shall briefly point out the similarities and differences in the Schr\"odinger and Heisenberg approaches. To simplify calculations without compromising physical insight, in the Schr\"odinger picture we restricted the maximum number of excitations in the atomic gas to one. In the Heisenberg picture, this constraint is relaxed, and the system of two nanocantilevers interacting with the ultracold BEC is effectively treated as three coupled harmonic oscillators. In reality, however, the number of excitations in the atomic gas is limited by the total number of atoms $N$. The Heisenberg solution is therefore valid as long as the total number of excitations is less than the total number of atoms $N$, so that the gas is effectively equivalent to a harmonic oscillator. The Schr\"odinger and Heisenberg approaches give identical results if this holds, and if more than one excitation is allowed in the gas in the Scr\"odinger picture.

\section{Dissipative dynamics} 
\label{sec:dissipation}
So far we have neglected decay channels in our system, but with any real physical system, dissipation is of course inevitable. Dissipation can occur in the atomic gas  with a decay rate $\Gamma$, or by thermal decay of the nanocantilevers  with a decay rate $\gamma$. 
For a nanocantilever with resonant frequency $\omega_{0}/2\pi= 1$ ${\rm MHz}$ and quality factor $Q=10^{6}$ \cite{NoteHighQ}, one obtains $\gamma \sim 1{\rm ~ Hz}$ \cite{roch}, 
whereas $\Gamma$ is largely governed by spin flips due to collisions or stray currents in the magnet or the cantilever \cite{magfield1,magfield2}. We have chosen $\Gamma =10$ ${\rm Hz}$ in the examples below, which allows us to neglect any direct cantilever dissipation and consider the much more rapid decay of the ultracold gas as the only dissipation channel. 
We note also that if the ultracold gas loses an excitation, then it is very unlikely that a cantilever can pick 
it up, owing to their very small solid angles as seen from the ultracold gas.

\begin{figure}[t]
   \begin{center}
	\includegraphics[width=0.5\textwidth]{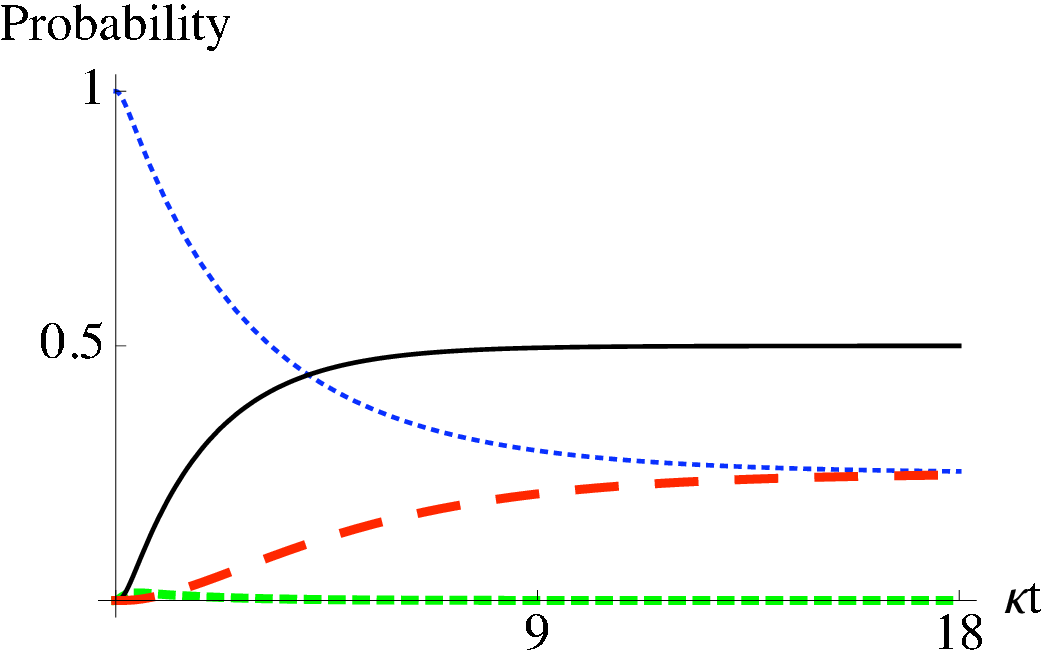}
        \caption{\label{fig4} (Color online) Time evolution of the  occupation probability for the states $|g,0,0\rangle$ (solid), 
        $|g,1,0\rangle$ (dotted),  $|g,0,1\rangle$ (thick dashed) and $|e,0,0\rangle$ (thin dashed). Initially $C_{g,1,0}(0)=1$.
        }
   \end{center}
\end{figure}

Under the Born-Markov approximation, evolution including dissipation  \cite{hpbr} is well described by a  Lindblad-type master equation of the form 
\begin{equation}\label{13}
\frac{\partial}{\partial t} \hat{\rho} =\frac{-i}{\hbar}[H,\hat {\rho}]+\mathcal L\hat{\rho} ,
\end{equation}
where $\hat{\rho}$ is  the  density matrix of the system, and 
\begin{equation}\label{14}
 \mathcal L\hat{\rho}    \equiv \frac{\Gamma}{2} (2\hat{J_{-}}\hat{\rho} \hat{J_{+}}-\hat{J_{+}}\hat{J_{-}}\hat{\rho}-\hat{\rho} \hat{J_{+}}\hat{J_{-}})
\end{equation}
is the Lindblad operator. 
With at most one excitation present in the system, Eq. \eqref{14}  can be solved  by evaluating 
 \begin{equation}
(\mathcal L\rho)_{i,j}= \langle i|\left(\sum_{k=0,l=0}^3 \mathcal L\rho_{kl}|k\rangle \langle l|\right)|j \rangle
\end{equation}
where $|0\rangle=|g,0,0\rangle$, $|1\rangle=|e,0,0\rangle$, $|2\rangle=|g,1,0\rangle$ and $|3\rangle=|g,0,1\rangle$. In this basis, Eq. \eqref{13} transforms to a set of coupled differential equations, 
 \begin{eqnarray}\label{a}
 \frac{\partial}{\partial t} \hat{\rho}+\frac{i}{\hbar}[H,\hat{\rho}]=
-\frac{N \Gamma}{2}\left[ \begin{array}{cccc} 
 {-2\rho_{1,1}}&\rho_{0,1}&{0} & {0} \\ 
 \rho_{1,0}&{2\rho_{1,1}}& \rho_{1,2}&\rho_{1,3}\\  
 {0}&\rho_{2,1}& {0}&{0}\\
 {0}&\rho_{3,1}& {0}&{0}\\
   \end{array}
\right]
\end{eqnarray}
A numerical solution of Eq. \eqref{a}, for the initial condition  $C_{g,1,0}(0)=1$, is shown in Fig.~\ref{fig4}.
In the steady state, the system relaxes to a statistical mixture of the ground state and the dark state with equal probability. Hence dissipation-assisted time evolution leads to a long-lived 
entangled state of two nanocantilevers. 

\begin{figure}[t]
   \begin{center}
	\includegraphics[width=0.5\textwidth]{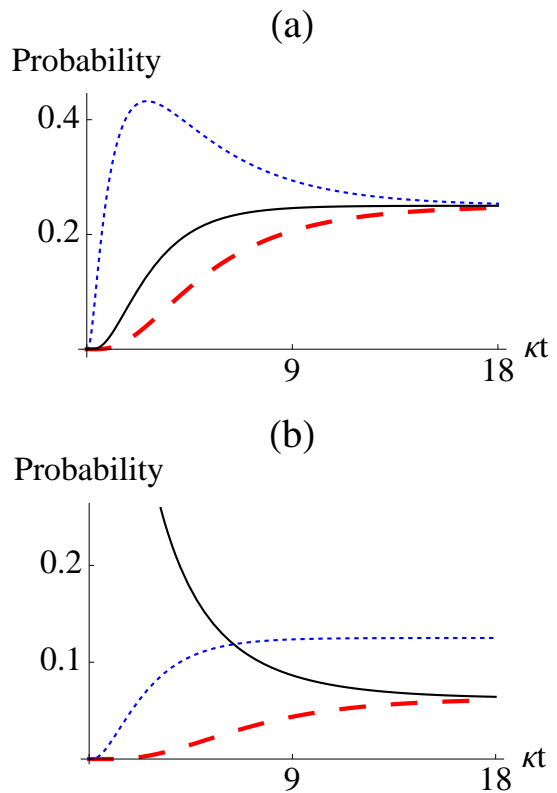}
        \caption{\label{fig5} (Color online)  Occupation probabilities, as a function of time, for the states (a) $|g,0,0\rangle$ (solid), $|g,0,1\rangle$ (dotted), and $|g,1,0\rangle$ (dashed); (b) $|g,0,2\rangle$ (solid), $|g,1,1\rangle$ (dotted), and $|g,2,0\rangle$ (dashed). Initially $C_{g,0,2}(0)=1.$
               Excitations in the ultracold gas decay so quickly that the probability  for states containing such excitations to be occupied are much smaller
        than the probabilities shown here.}
   \end{center}
\end{figure}

Following a similar series of steps as outlined above, we have solved for the dissipative dynamics with two and three excitations initially present in the system,
and have found similar behavior.  For an initial state  where  $C_{g,2,0}(0)=1$, a numerical solution of Eq. \eqref{13} 
is shown in fig.~\ref{fig5}. 
The  system relaxes to a statistical mixture of dark states with zero, one, and two excitations. With the chosen initial state, the probabilities for these dark states to be occupied are $0.25, 0.5$ and $0.25$ respectively. For three excitations in the initial state (not shown here), the steady state is a mixture of dark states with zero, one, two and three excitations, with respective statistical weights of $0.125, 0.375, 0.375$ and $0.125$ for an initial state $C_{g,3,0}(0)=1$. 

\begin{figure}[t]
   \begin{center}
   \includegraphics[width=0.5\textwidth]{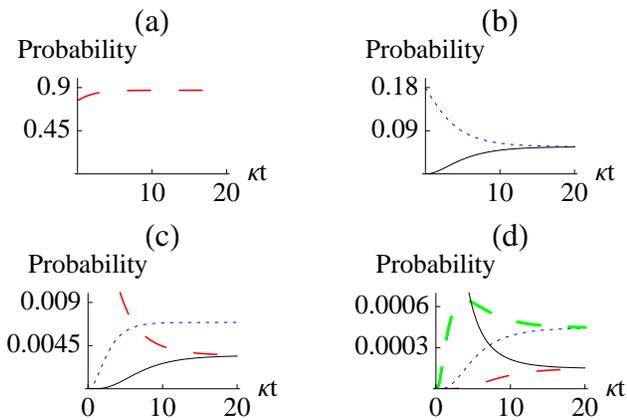}
\caption{\label{fig6} (Color online)  Occupation probabilities, as a function of time, for the states (a) $|g,0,0\rangle$ (dashed), 
(b) $|g,1,0\rangle$ (dotted),  $|g,0,1\rangle$ (solid)
, (c) $|g,2,0\rangle$ (dashed), $|g,1,1\rangle$ (dotted),  $|g,0,2\rangle$ (solid)
, (d) $|g,3,0\rangle$ (solid), $|g,2,1\rangle$ (thick dashed),  $|g,1,2\rangle$ (dotted) and $|g,0,3\rangle$ (thin dashed).
Cantilever `a' is initially in a mixed state of zero, one, two and three excitations, approximating a thermal state with $\langle n_{\it thermal}\rangle=0.3$.  
Excitations in the ultracold gas decay so quickly that the probability  for states containing such excitations to be occupied
are much smaller than for the other states considered here.}
\end{center}
\end{figure}

The robustness of the scheme lies in the fact that entangled states of nanomechanical systems may be generated even if these 
are initially found in thermal states. This will be further investigated in future work. Here, we study a simple example. We solve for the dynamics of the system with one of the cantilevers and the ultracold gas prepared in their respective ground states. The other cantilever is in a mixture of zero, one, two and three excitations, with a distribution that approximates that of a thermal state. The solution of the master equation is shown in fig.~\ref{fig6}, and the resulting steady state is entangled. To summarize this section, we have shown that  two nanocantilevers, interacting with 
 the same dissipative cloud of 
ultracold atoms, will become entangled with each other. The entangled state is immune to dissipation in the ultracold gas, and hence opens the possibility of achieving long-lived entangled states of nanomechanical systems. 

\section{Entanglement measures}
\label{sec:entanglement}
After showing that it is possible to generate long-lived entangled states of two nanocantilevers, we will now quantify the entanglement between these. In section \ref{sec:unitarysch}, we used the Schr\"odinger equation to study the unitary evolution of the system of two cantilevers interacting with an ultracold BEC with one, two and three excitations present in the system. The time evolved state of the two nanocantilevers is  a mixed state found by tracing over the atomic gas. 

To quantify the entanglement in a bipartite system 
in a mixed state, we use the Peres criterion \cite{pere}. We compute the negativity for the reduced density matrix, defined as
\begin{equation}
\label{negativity}
\mathcal N = max(0,- \sum_{i}\lambda_{i})
\end{equation}
where $\sum_{i}\lambda_{i}$ is the sum of  all the negative eigenvalues of the partially transposed reduced density matrix \cite{jlee}.
The negativity $\mathcal N$ is not only easy to compute but has an added advantage of  being an entanglement monotone \cite{gvid}. 

\begin{figure}[t]
   \begin{center}
	\includegraphics[width=0.5\textwidth]{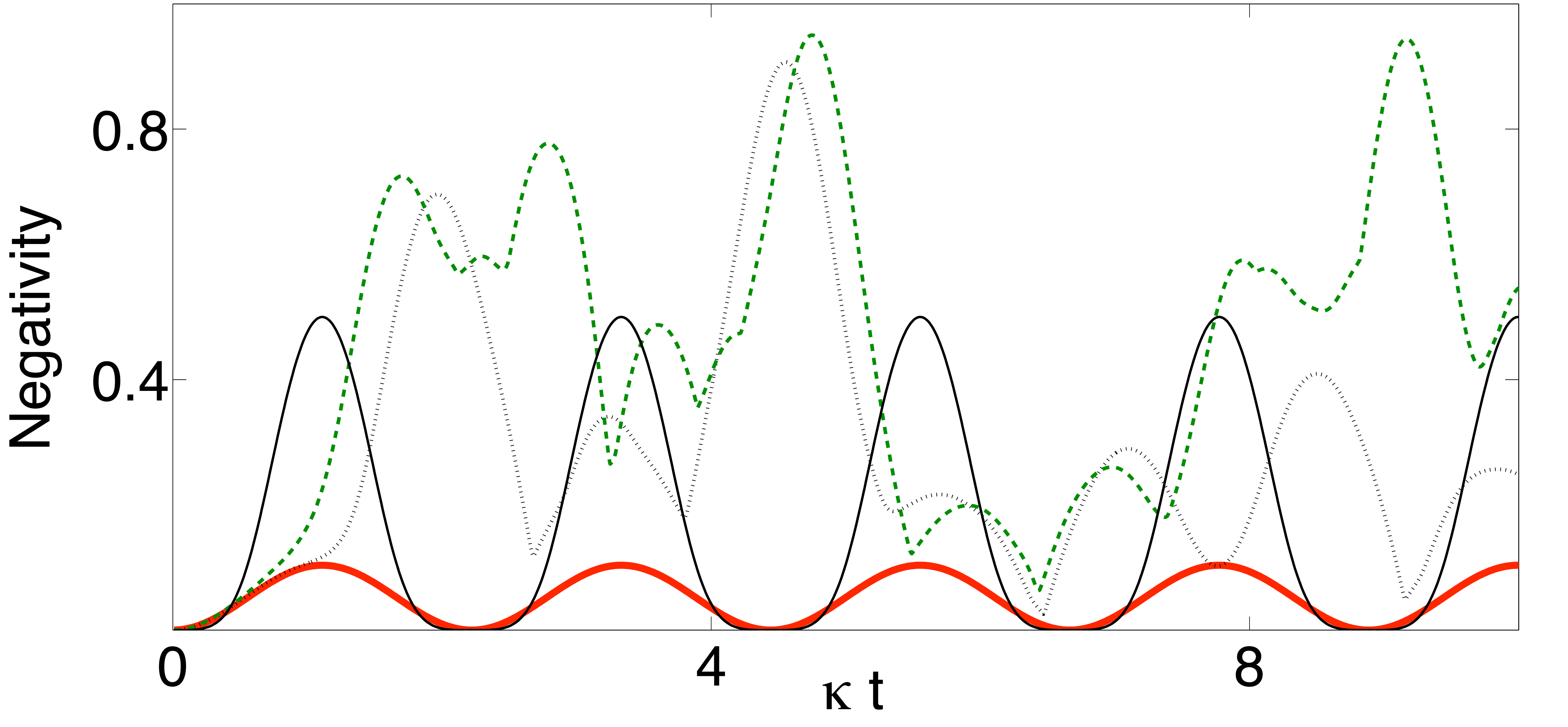}
        \caption{\label{fig7} (Color online)
       Degree of entanglement, as measured by the negativity defined in Eq.~(\ref{negativity}),
        for a system of two nanocantilevers interacting with a dissipation free ultracold gas, in the one-ecxitation (thick solid), two-excitation (dotted) and three-excitation (dashed) subspaces, with all the excitations initially present in one of the cantilevers. Also, for  comparison, the negativity is presented for the case when the initial excitation is in the gas for the one-excitation subspace (thin solid). 
        }
   \end{center}
\end{figure}

\begin{figure}[t]
   \begin{center}
	\includegraphics[width=0.5\textwidth]{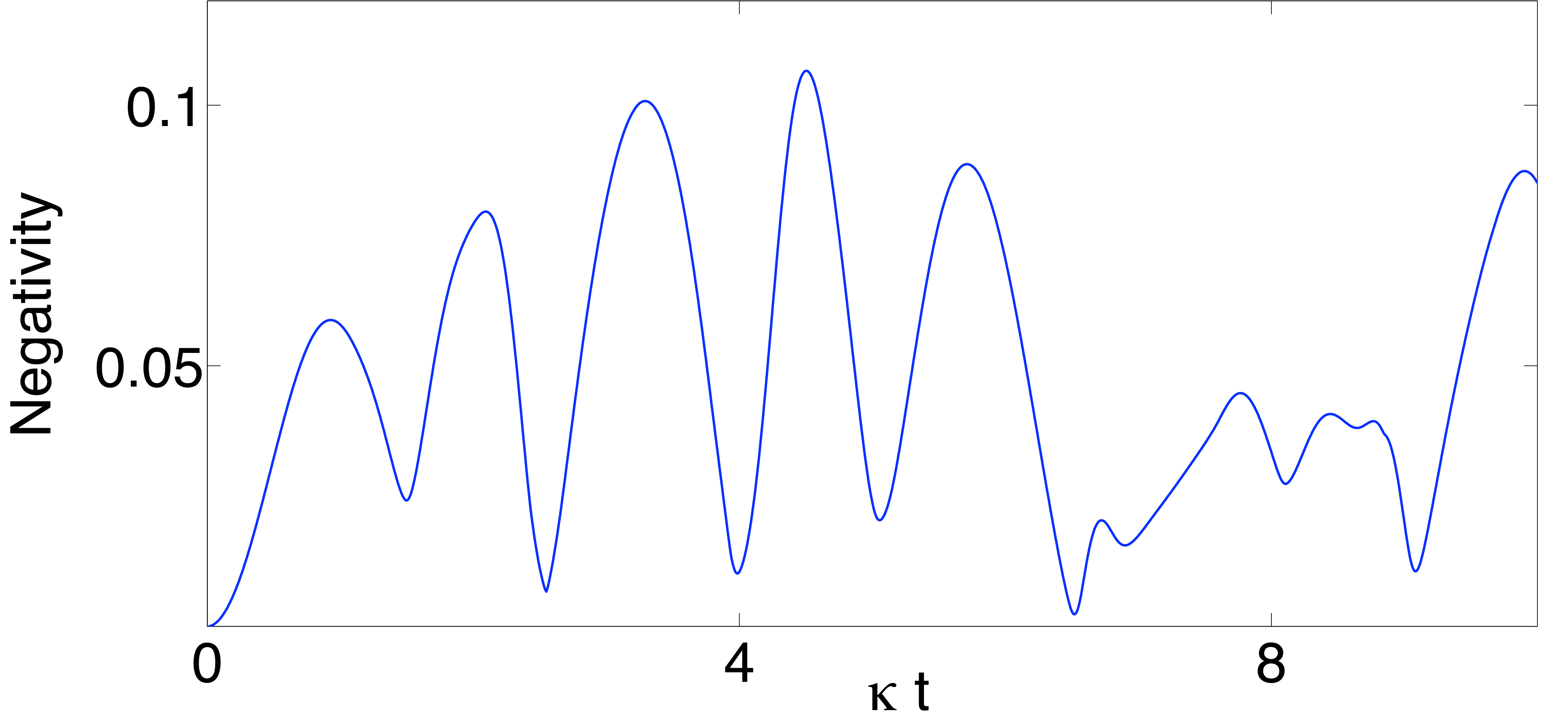}
        \caption{\label{fig8} (Color online)
       Degree of entanglement, as measured by the negativity 
        for a system of two nanocantilevers interacting with a dissipation free ultracold gas for an initial mixed state of the first three excitation subspaces with average thermal occupancy of 0.3.}
   \end{center}
\end{figure}

Using the time evolved wave functions obtained in \ref{sec:unitarysch},  the negativity $\mathcal N$, as defined in Eq.~(\ref{negativity}), is computed and is shown in fig.~\ref{fig7}. For a bipartite system of two qubits, $\mathcal N$ lies between 0 and 0.5, with $\mathcal N=0$ for separable states and $\mathcal N=0.5$ for maximally entangled states. For a pure state of two qudits, with $k$ terms in its Schmidt decomposition,  $0\leq \mathcal N \leq (k-1)/2 $. As shown in fig.~\ref{fig7}, in the one-excitation subspace the two nanocantilevers are entangled at all times, except at certain instants when the negativity becomes equal to zero. Interestingly,  in higher excitation subspaces, the two nanocantilevers are always entangled for $t > 0$. In fig.~\ref{fig7} we have also compared $\mathcal N$ for two different initial conditions in the one-excitation subspace. As is clear from fig.~\ref{fig7}, if the initial excitation lies in the ultracold gas, then maximally entangled states of the two nanocantilevers occur at certain times. One  of the main features of our proposed scheme is 
that entangled states of the two nanocantilevers are generated even for an initial classical state. 
As an illustration, we plot in fig.~\ref{fig8} the negativity as a measure of entanglement for an initial separable mixture of the first three excitation subspaces. A non zero value of negativity in fig.~\ref{fig8} again points to a finite entanglement being present between the two nanocantilevers.


As discussed in the previous section, in the presence of dissipation, a system of two nanocantilevers coupled to an ultracold gas of atoms relaxes to a statistical mixture of various dark states. The dark states are pure entangled states of the two nanocantilevers. For pure states, it is possible to quantify entanglement using the \emph{participation ratio} $\zeta$ \cite{kwch}, defined as
\begin{equation}
\zeta=\frac{1}{\sum_{i=1}^{k} p_{i}^{2}},
\end{equation}
where $p_{i}$ are the coefficients  in the Schmidt decomposition of the pure state $|\Psi\rangle$ of the bipartite system.
The quantity $\zeta$ lies between $1$ for separable states and $k$ for maximally entangled states. For the dark states in the one-, two- and three-excitation subspaces we have $\zeta_{1,2,3} =2, 2.67$ and $3.2$ respectively. Although $\zeta_{3} >\zeta_{2} > \zeta_{1}$, only the dark state in the one-excitation subspace is a maximally entangled state of the two nanocantilevers. Nonetheless, in higher excitation subspaces there is a finite degree of entanglement present between the two nanocantilevers. 

\section{Conclusions} 
\label{sec:conclude}
We have proposed a scheme to generate entanglement between two nanomechanical systems through indirect interactions mediated by an ultra cold gas of atoms. The excitation energy of the two-level atoms is assumed to have been tuned to coincide with the energy quantum of the fundamental flexural mode of the two identical cantilevers. Hence single excitations have the same energy in the atomic system as in the cantilevers and we have studied the dynamics of systems with a total of one, two and three excitations present. In the presence of dissipation in the atomic gas, the two nanocantilevers 
relax into a statistical mixture of various dark states of the system.  This opens the possibility to achieve long-lived entangled states of nanomechanical systems which are only limited by the dephasing time of the cantilevers. 
The atomic decay channel is important for the creation of time independent entangled states, 
provided that the decay rate of the atoms is low enough to enable a strongly coupled regime and high enough 
so that the steady state is reached before the coherence of the cantilevers decays away. 

With respect to the preparation of an initial state it has been shown in  \cite{ckla}  that if a two level system  is coupled to a classical field  and a quantum  field, prepared in its vacuum state,  and if the two coupling strengths can be tuned independently of each other  then atom-field interactions can force the vacuum state of the field to an arbitrary superposition of the Fock states. This can be used for the initial state preparation of the nanocantilevers.
Also, in a recent work \cite{fari} a possibility of transferring also non-gaussian states from an optical field to a mechanical oscillator has been outlined.

A challenging aspect of any scheme involving entanglement generation between 
``macroscopic'' mechanical systems is the actual  experimental detection of entanglement.  
A promising scheme suggested in \cite{mdla} employs a quantum non demolition (QND) 
measurement of the excitation of a mechanical oscillator by using the anharmonic coupling 
between two beam bending modes,  which uniquely shifts the resonant frequency of the 
readout oscillator. Another recent  proposal is to determine the Wigner function of a mechanical 
resonator by coupling the vibrational mode of the oscillator to a three level ``detector'' atom, 
which in turn is coupled to a pair of optical fields, inducing Raman transitions between its 
ground and excited state \cite{ssin}.  It has been shown that the probability for the ``detector'' 
atom to be found in the ground state  is a direct measure of the Wigner characteristic function 
of the nanomechanical oscillator as long as dissipation in the system can be kept under control. 
Entanglement detection for nanomechanical systems is thus experimentally challenging, but 
may be possible using techniques such as these.

\section{Acknowledgements} We gratefully acknowledge useful discussions with Aidan Arnold, Stephen Barnett, Michael Hall and Mark Hillery. C. J. acknowledges the ORS scheme, and M. J. partial support from the Korean WCU program funded by MEST through the NFR (Grant No. R31-2008-000-10057-0).


\begin{thebibliography}{99}
\bibitem{altgeomtry} 
A trap geometry similar to an Ioffe-Pritchard trap would also be possible, in which case the ferromagnets would be aligned along the $z$-direction and $B_0$ along the long trap axis in the $x$-direction.
\bibitem{FiniteMagnetRemark}
If the finite size of the magnet is taken into account the coupling constant $\kappa$ is reduced by a factor $\sim2$. The strong coupling regime can still be achieved by increasing the number of atoms in the trap center.
\bibitem{NoteBroadening}
For a large BEC with strong collisional interactions there will be inhomogeneous broadening present, hence all atoms will not necessarily precess with a single Larmor frequency. The result is a reduction of the effective number of resonantly coupled atoms. This practical hindrance can be overcome by 
controlling the collisions in the gas by means of adjusting the scattering length using techniques like Feshbach 
resonances, or using a lower density of atoms.
\bibitem{NoteHighQ}
Such high Q-values have so far only been achieved with doubly clamped prestressed resonators \cite{roch}. 
We note that our analysis could equally well be applied to a setup where two such resonators interact via a BEC.
\end{thebibliography}
\end{document}